\documentclass[PAPER, cernpreprint=true, atlasdraft=false, UKenglish, texmf, orcidlogo]{atlasdoc}
\usepackage[default,subcaption=true,multirow=true]{atlaspackage}
\usepackage{atlasbiblatex}

\usepackage[BSM=true,hepprocess=true]{atlasphysics}
\usepackage{pifont}
\usepackage{subcaption}
\graphicspath{{logos/}{figures/}}

\usepackage{ANA-HIGP-2024-01-PAPER-defs}

\AtlasTitle{Evidence of Higgs boson inclusive production at high transverse momentum decaying to a pair of $b$-quarks with the ATLAS detector}

\AtlasVersion{1.6}

\AtlasAbstract{%
This letter reports on the first evidence of inclusive Higgs-boson production at high transverse momentum in the $b\bar{b}$ final state, reconstructed in a single large-radius jet. The results are based on proton proton collision data recorded by the ATLAS detector at the Large Hadron Collider at a centre-of-mass energies of 13 TeV and 13.6 TeV, corresponding to a total integrated luminosity of 301 fb$^{-1}$. The study profits from the large background suppression provided by the use of a new transformer-based algorithm for jet identification and the sharper mass and transverse momentum, $p_{\text{T}}$, resolution from a dedicated regression model. The yield relative to the Standard Model prediction, for Higgs bosons produced at $p_{\text{T}}$ larger than 450 GeV is measured to be $1.53 \pm 0.27 \, ^{+0.33}_{-0.27}\pm 0.17$, where the first uncertainty is statistical, the second is experimental systematic, and the third accounts for theory systematic, corresponding to an observed (expected) significance of $3.8\sigma(2.5\sigma)$ relative to the background-only hypothesis. Results are also obtained in three Higgs boson $p_{\text{T}}$ intervals and found to be compatible with Standard Model predictions.
}

\AtlasRefCode{HIGP-2024-01}

\PreprintIdNumber{CERN-EP-2026-082}

\AtlasJournal{PRL}
\AtlasJournalRef{Phys. Rev. Lett. 137, 071801}
\AtlasDOI{https://doi.org/10.1103/vbxg-71c2}

\AtlasCoverEgroupAnalysisTeam{atlas-higp-2024-01-analysis-team@cern.ch}

\hypersetup{pdftitle={ATLAS document},pdfauthor={The ATLAS Collaboration}}

\begin{document}

\maketitle

%

%
%

% The next lines are included from the .//sections/1.introduction.tex input file
%
%

The steady increase in the amount of data collected in high-energy proton--proton ($pp$) collisions at the CERN Large Hadron Collider (LHC), together with advances in reconstruction techniques using transformer-based~\cite{VaswaniAttention,ATL-PHYS-PUB-2023-021,ATL-PHYS-PUB-2024-015} models, opens new avenues for exploring phenomena at the TeV energy scale.
Although measured production and decay rates are so far consistent with the Standard Model (SM)~\cite{HIGG-2021-23,CMS-HIG-22-001}, the study of Higgs-boson production at large transverse momentum (\pT) offers the opportunity to unveil or constrain new phenomena~\cite{Grojean:2013nya, Dawson:2015gka, Grazzini:2016paz, Banfi:2018pki}. At high \pT, Higgs boson production becomes sensitive to the loop structure of gluon--gluon fusion (\ggF). As a result of modified couplings from possible new physics contributions, effects from physics beyond the SM are typically enhanced by factors depending on the Higgs boson \pT. This motivates a dedicated study in the boosted regime with \pT\ up to and beyond 1 TeV. The \Hbb decay dominates the branching fractions and offers the best sensitivity of this rare process~\cite{CMS-HIG-17-010,HIGG-2018-52,HIGG-2021-08,HIGG-2021-11,CMS-HIG-21-020}.

This letter reports a study of boosted inclusive Higgs boson production in $pp$ collisions recorded by the ATLAS detector.
Signal events feature a central high-\pT jet containing the two collimated $b$-hadrons from a \Hbb decay, recoiling against additional jets.
Because of energy lost through neutrinos from $b$-hadron decays and of detector resolution effects, the Higgs boson candidate may be reconstructed as either the leading or subleading jet in \pT.

The Higgs boson yield is extracted from the jet mass \mJ distribution, dominated by a smoothly falling non-resonant multijet QCD background with additional contributions from hadronic \Vjets and top-quark pair (\ttbar) production.
The resonant \Zbb and \Hbb rates are expressed in terms of signal-strength modifiers $\mu$, defined as the ratio of the measured yields to their SM predictions, in a simultaneous fit that determines the multijet background entirely from data.
The pronounced \Zbb peak serves as an in-situ calibration of the jet tagging rates and the jet mass scale and resolution in the fit used to extract the Higgs boson signal.

Compared to an earlier Run~2 study~\cite{HIGG-2021-08}, the analysis sensitivity improves by a factor of seven to ten.
This improvement results from stronger multijet background suppression with a new transformer-based $bb$ tagger, which also nearly removes the \Wqq and \ttbar contributions from the signal region, leaving \Zbb and \Hbb as the dominant resonant processes.
Additional gains arise from the improved jet mass and \pT resolution from a transformer-based regression model, the simultaneous fit of signal and control regions to constrain the multijet background, and a larger analysed dataset.

% End of text imported from the .//sections/1.introduction.tex input file

% The next lines are included from the .//sections/2.detector.tex input file
The ATLAS experiment at the LHC is a multi-purpose particle detector with a forward-backward symmetric cylindrical geometry, described in detail in Refs.~\cite{PERF-2007-01,GENR-2019-02}. It consists of an inner tracking detector (ID) surrounded by a thin superconducting solenoid, electro-magnetic and hadronic calorimeters, and a muon spectrometer.
This analysis uses $pp$ collision datasets recorded with the ATLAS detector totaling \SI{140}{\fb^{-1}} at $\sqrt{s}=13~\TeV$ during Run~2 (2015--2018) and \SI{161}{\fb^{-1}} at $\sqrt{s}=13.6~\TeV$ during the first three years of Run~3 (2022--2024)~\cite{DAPR-2021-01,ATL-DAPR-PUB-2024-001,ATL-DAPR-PUB-2025-001}.

Monte Carlo (MC) simulated event samples are used to model Higgs boson production at next-to-leading-order (NLO) QCD accuracy using \POWHEGBOX~\cite{Alioli:2010xd,Nason:2004rx,Frixione:2007vw} with parton showering and hadronization provided by \PYTHIA[8]~\cite{Sjostrand:2014zea}. Cross sections are normalized to state-of-the-art predictions, corresponding to NNLO QCD calculations for \ggF\ including exact NLO top-mass effects, NNLO QCD for vector-boson-fusion (\VBF), and NLO QCD for associated production with a vector boson (\VH) or top-quarks (\ttH)~\cite{Becker:2020rjp}.
The \Vjets and \ttbar backgrounds are simulated using \SHERPA[2.2.14]~\cite{Bothmann:2019yzt} and \POWHEGBOX~\cite{Frixione:2007nw,Nason:2004rx,Frixione:2007vw,Alioli:2010xd}, respectively. The detector response is modeled with \GEANT~\cite{Agostinelli:2002hh}, including pile-up effects~\cite{ATL-PHYS-PUB-2016-017}. The effects of radiation damage in the Pixel detector~\cite{IDET-2017-10,PIXE-2023-02} are included for samples matching Run~3 data conditions.

% End of text imported from the .//sections/2.detector.tex input file

% The next lines are included from the .//sections/3.reconstruction.tex input file
Charged-particle tracks are reconstructed in the ID~\cite{IDTR-2022-04,ATL-PHYS-PUB-2019-015} and used as components of the jet objects and for $b$-tagging.
Jets are reconstructed from unified flow objects (UFO)~\cite{JETM-2018-06} using the \antikt algorithm~\cite{Cacciari:2008gp} with $R=1.0$ and groomed with Soft-Drop~\cite{Larkoski:2014wba,Dasgupta:2013ihk} and pile-up mitigation techniques~\cite{Berta:2014eza,Cacciari:2014gra}.

Transformer-based models are a key aspect of the analysis and used to identify and calibrate jets from resonances decaying to \bbbar ($bb$-jet). These models are trained on samples of simulated events containing mixtures of jet flavors and exploit the kinematic and topological properties of UFO constituents and other charged particle tracks.

Jets containing two \bhadrons are identified using an improved version of the GN2X tagger~\cite{ATL-PHYS-PUB-2023-021} trained on simulated event samples corresponding to Run~2 and Run~3 conditions.
The GN2X tagger output is rescaled in simulated Run~2 events to account for pixel radiation damage.
Mass-dependent selections on the tagger output (working points, WPs) are defined independently for the leading and subleading jet \textit{categories}, \pT ranges, and data-taking years using multijet events simulated with \PYTHIA[8]~\cite{ATL-PHYS-PUB-2019-017}.
These WPs suppress the multijet background by a nearly constant factor of about 500 across the jet-mass range, an improvement of more than four times compared to that of Ref.~\cite{HIGG-2021-08}, with signal efficiencies of $\sim45$\% ($\sim40$\%) for leading (subleading) Higgs jets.
The efficiency for selecting \Zbb or \Hbb jets is similar, and the suppression of top-quark jets is comparable to that of multijet background jets.

The reconstructed jet mass and \pT\ response~\cite{JETM-2018-02} are improved using the bJR regression model~\cite{ATL-PHYS-PUB-2024-015}, enhancing the jet mass and \pT\ resolution by approximately 30\% and 20\%, respectively, and thereby increasing the analysis sensitivity by nearly 30\%.
The \pT\ scale and resolution are validated using \Zmm+jet events by exploiting the balance between the recoil jet and the $\mu^{+}\mu^{-}$ system.
Data and simulation agree at the 3\% level, defining the associated energy scale (JES) and resolution (JER) uncertainty.
Additionally, if a muon is associated to a subleading jet passing the GN2X requirement, its four-momentum is added to the jet, improving its mass resolution by about 12\% in simulated Higgs boson events.

The simulation-to-data $bb$-tagging efficiency correction factors \KGNTX, jet mass scale (JMS), and mass resolution (JMR) are determined in situ from the \Zbb peak and applied to the Higgs boson signal templates.
The factor $\KGNTX \equiv \muZbb/\muZmm$, where \muZmm is obtained from an independent fit to \Zmmjets data, ensures that common \Zjets production uncertainties versus jet \pT cancel in the ratio~\cite{ATL-PHYS-PUB-2021-035}.

Data events are selected using a combination of unprescaled high-level triggers requiring a jet with online \pT thresholds from 420~GeV, with additional jet mass requirements used from 2018~\cite{TRIG-2019-04,TRIG-2022-01}.
Events are required to contain at least one \textit{candidate} jet spatially matched to the jet in the trigger system, pseudorapidity $|\eta|<2.0$, transverse momentum on the trigger efficiency plateau of $\pT>450~\GeV$, \mJ in excess of \SI{45}{\GeV} and satisfying $2\mJ/\pT < 1$, ensuring well-contained boosted topologies. After applying the jet corrections above, the study focuses on the jet mass range \SIrange{55}{160}{\GeV}.

Candidate jets are organized into \textit{regions} according to the GN2X tagger output.
If the leading jet passes the GN2X requirement, it enters the leading-jet signal region (SRL).
Otherwise, if the subleading jet is tagged, it enters the subleading-jet signal region (SRS).
Jets failing the tagger requirement have an \mJ distribution that closely resemble that of the SR due to the definition of the GN2X WPs.
These jets are partitioned into a control region (CR), used to constrain the multijet background shape in the SR, and a validation region (VR) used to test the multijet model.
The CR is selected to have ten times the SR yield, proportionally across data-taking periods to preserve the time-dependent detector and pile-up conditions; the remaining 97.5\% of anti-tagged jets form the VR.
Anti-tagged leading (subleading) candidate jets populate the CRL and VRL (CRS and VRS).
All regions are studied in three candidate-jet \pT\ bins: \SIrange{450}{650}{\GeV}, \SIrange{650}{1000}{\GeV}, and above \SI{1000}{\GeV}.

Within the SR, the SM predicts that \ggF\ production constitutes about 50\% of the selected signal in the \SIrange{450}{650}{GeV} \pT interval, decreasing to $\sim$40\% above \SI{1}{TeV}, while the \VH fraction increases from $\sim$25\% to nearly 40\%, whereas VBF and \ttH remain subleading at the $\sim$20\% and $\sim$5\% level, respectively.

For the bJR validation, trigger efficiency studies,
and for an input to \KGNTX, the \Zmmjets sample is selected by requiring two oppositely charged muons with $\pT>50~\GeV$ and a high-\pT\ recoiling jet, using the same jet trigger requirements discussed above.
The \Zmm signal rate is measured in the same three \pT intervals considering a full suite of systematic uncertainties. The jet trigger efficiency is studied using \Zmm events collected by a muon trigger~\cite{TRIG-2018-01}.

% End of text imported from the .//sections/3.reconstruction.tex input file

% The next lines are included from the .//sections/4.modelling.tex input file
The \mJ spectrum of the multijet background in each SR and the corresponding CR is described by an exponentiated polynomial of order $N$ ($f_N$), while residual SR--CR shape differences are absorbed by a multiplicative transfer function of order $M$ ($t_M$) multiplied to the CR \mJ spectrum,
\[
f_N(x|\vec{\phi}) = \phi_0 \exp\!\left(\sum_{i=1}^N \phi_i x^i\right), \ \ t_M(x|\vec{\psi}) = 1 + \sum_{j=1}^M \psi_j x^j
\]
where $x = (\mJ - 140~\GeV)/70~\GeV$, $\phi_i$ and $\psi_j$ are free parameters fitted independently to data for each jet category, \pt bin, and dataset. The $\phi_0$ normalizations are treated independently in the SR and CR.
Studies conducted with a set of MC generators and fragmentation models indicate that SR--CR \mJ\ shape variations can arise from differences in the flavor composition and jet substructure and can be well described by the transfer function. More details are given in the Appendix.
Compared to fits to the SRs alone, the use of the CRs helps to constrain the multijet background model. The uncertainties on the $f_N$ parameters are reduced by \SIrange{20}{30}{\%}, improving the statistical precision on \muH by $\simeq$15--20\%. The correlations between \muH and the dominant systematic uncertainties, JMS and JMR, are also reduced by \SIrange{15}{40}{\%}, leading to smaller systematic uncertainties.

The VR mass spectrum provides a good approximation of the SR multijet background. The determination of the multijet background is studied using VR events divided into statistically independent subsets, or \textit{slices}, with event counts comparable to the SR.
After subtracting the \Wboson, \Zboson and \ttbar resonant contributions, these slices are reweighted to match the SR multijet shape using the ratio of $f_N$ functions obtained from independent fits to the SR and CR spectra. \Zbb and \Hbb templates are then injected at varying signal rates to evaluate the linearity of the fit response and the presence of spurious signal effects~\cite{ATL-PHYS-PUB-2020-028}.
An ensemble of these slices
are used to perform fits with $3 \leq N \leq 5$ and $0 \leq M \leq 2$ for the multijet parametrization.

The $N$ and $M$ orders are determined using these fits~\cite{HIGG-2021-08}.
They are chosen such that the average bias of the extracted signal strength is small and the fraction of fits in which $\mu$ deviates by more than twice the fit statistical uncertainty \sigmafit from the mean is approximately 5\%. In each region, these studies show that $N=4$ and $M=2$ provide stable behavior.
Small residual biases of 0.1--0.3\,\sigmafit in the extracted Higgs boson signal strength are incorporated as a spurious-signal uncertainty, with a minor effect on the total uncertainty.
For the \Zbb rate, the residual bias is $\sim0.15\sigmafit$.

The strong rejection of more than two-prong jets by the GN2X tagger and the \mJ$<$160~GeV requirement greatly reduce the top-quark background.
A boosted semi-leptonic \ttbar control region is used to extract simulation-to-data correction factors from \mJ fits and are applied separately in the SRs and CRs with a 25\% uncertainty~\cite{TOPQ-2022-24}, leading to a 5\% uncertainty on \muH.

% End of text imported from the .//sections/4.modelling.tex input file

% The next lines are included from the .//sections/5.results.tex input file
Signal yields are extracted using a simultaneous profile likelihood fit~\cite{RooFit,Gligorov2021Avoiding}
to the jet mass distributions in the SRL, SRS, CRL, and CRS categories divided into the three jet \pT\ bins separately for the Run~2 and Run~3 datasets. In total, 12 pairs of signal and control regions are included in the fit.
The \Zbb event rates in the \KGNTX factors are treated as independent free parameters for each jet category, \pT\ bin, and dataset.
The likelihood depends on \Hbb signal strength parameters $\vec{\muH}$, 12 \KGNTX factors, over 100 systematic uncertainty parameters, and 96 free parameters $\vec{\phi}$ and $\vec{\psi}$ modeling the multijet background.

Higgs boson cross section uncertainties are taken from Ref.~\cite{Becker:2020rjp}, with additional top-quark-mass scheme effects from Ref.~\cite{ATL-PHYS-PUB-2023-031}. Acceptance uncertainties (scale, parton distribution function (PDF), and parton shower variations) are evaluated from simulation, yielding an uncertainty of \SIrange{17}{18}{\%} depending on \pT. These uncertainties are fully correlated across regions and are the only systematic uncertainties correlated between datasets.

The dominant experimental uncertainties arise from the JMS and JMR, shared between the \Zbb and \Hbb templates within a region but treated independently across jet categories, \pT bins, and datasets. Conservative prior uncertainties of 5\% (JMS) and 25\% (JMR) are assigned and are strongly constrained by the data: the \Zbb peak position is determined to better than 1\% precision in the lowest \pT bin and about 2\% at the highest \pT, while the mass resolution is constrained to \SIrange{3}{18}{\%} depending on \pT. These constraints are propagated to the \Hbb template. Separate JMS and JMR parameters are used for the \Vqq contribution in the CRs to account for flavor-composition differences. The full list of systematic uncertainties is given in the Appendix.

\begin{figure}[htb!]
\centering
\includegraphics[width=.5\columnwidth]{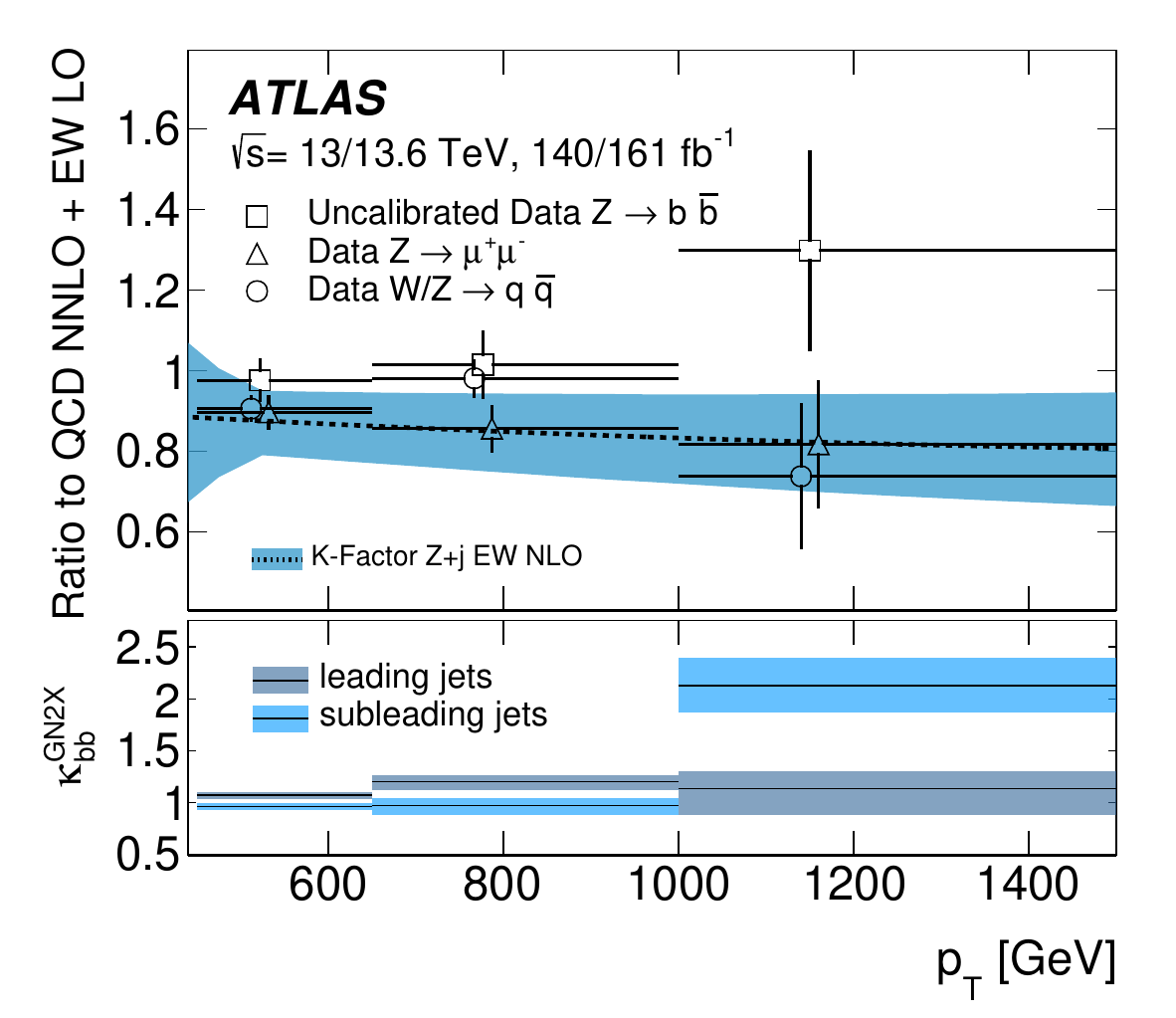}
\caption{Observed \Zmmjets event rate as a function of jet \pT compared to theoretical predictions from \SHERPA~2.2.14 including QCD NNLO corrections from NNLOJET~\cite{Ridder:2015dxa,Ridder:2016nkl} and electroweak (EW) NLO corrections from \SHERPA with uncertainties estimates from Ref.~\cite{Lindert:2017olm}. The \Vqq and \Zbb rates, the latter uncalibrated for GN2X efficiency, are overlaid for comparison (upper panel). Fitted values of the leading and subleading \KGNTX factors averaged over datasets are shown within each \pT bin in the lower panel.
\label{fig:ZvsJetPt}}
\end{figure}

\begin{figure}[t]
\centering
\begin{subfigure}{0.32\columnwidth}
\includegraphics[width=\textwidth]{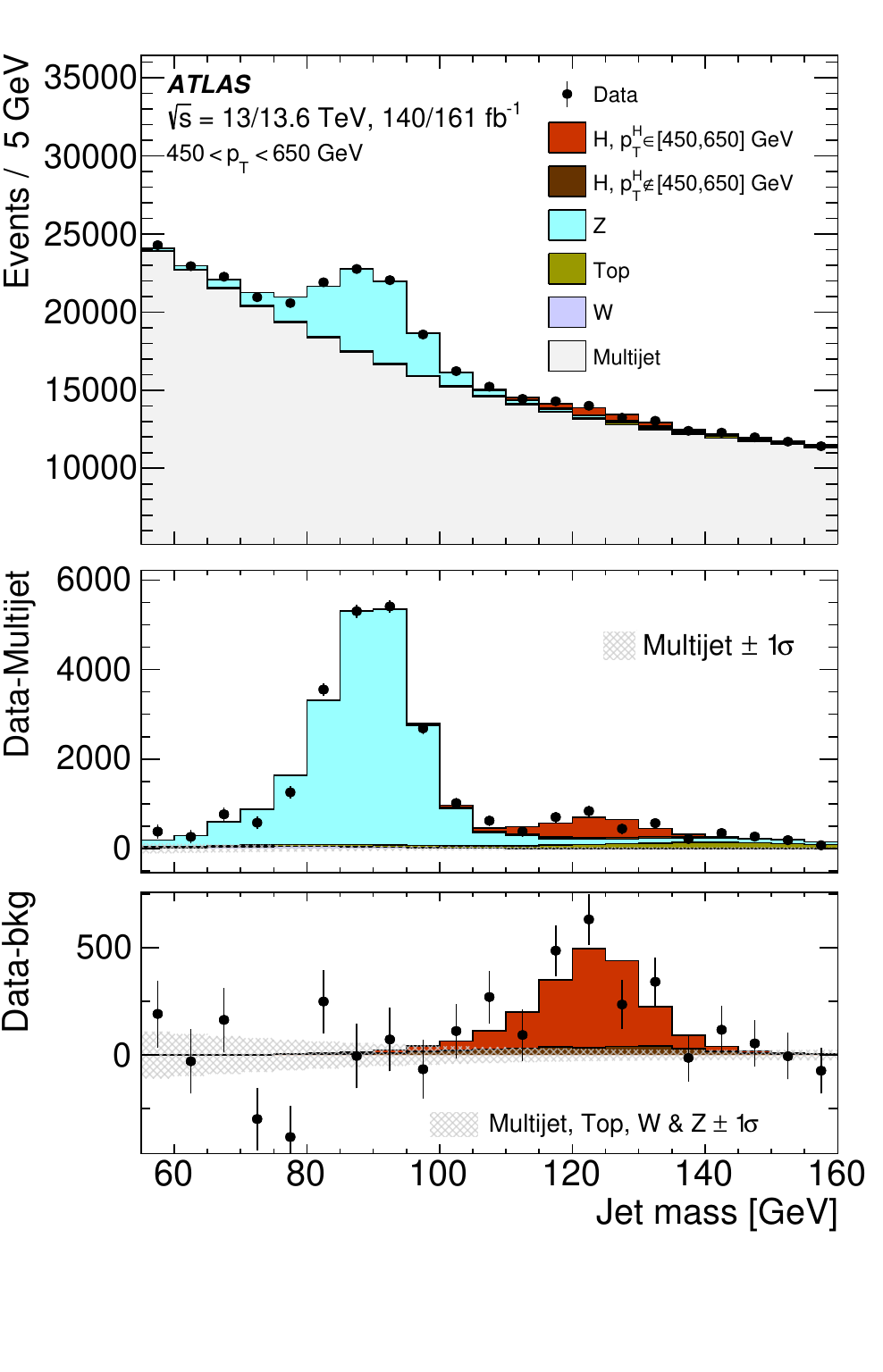}
\vspace*{-1cm}
\caption{}
\end{subfigure}
\begin{subfigure}{0.32\columnwidth}
\includegraphics[width=\textwidth]{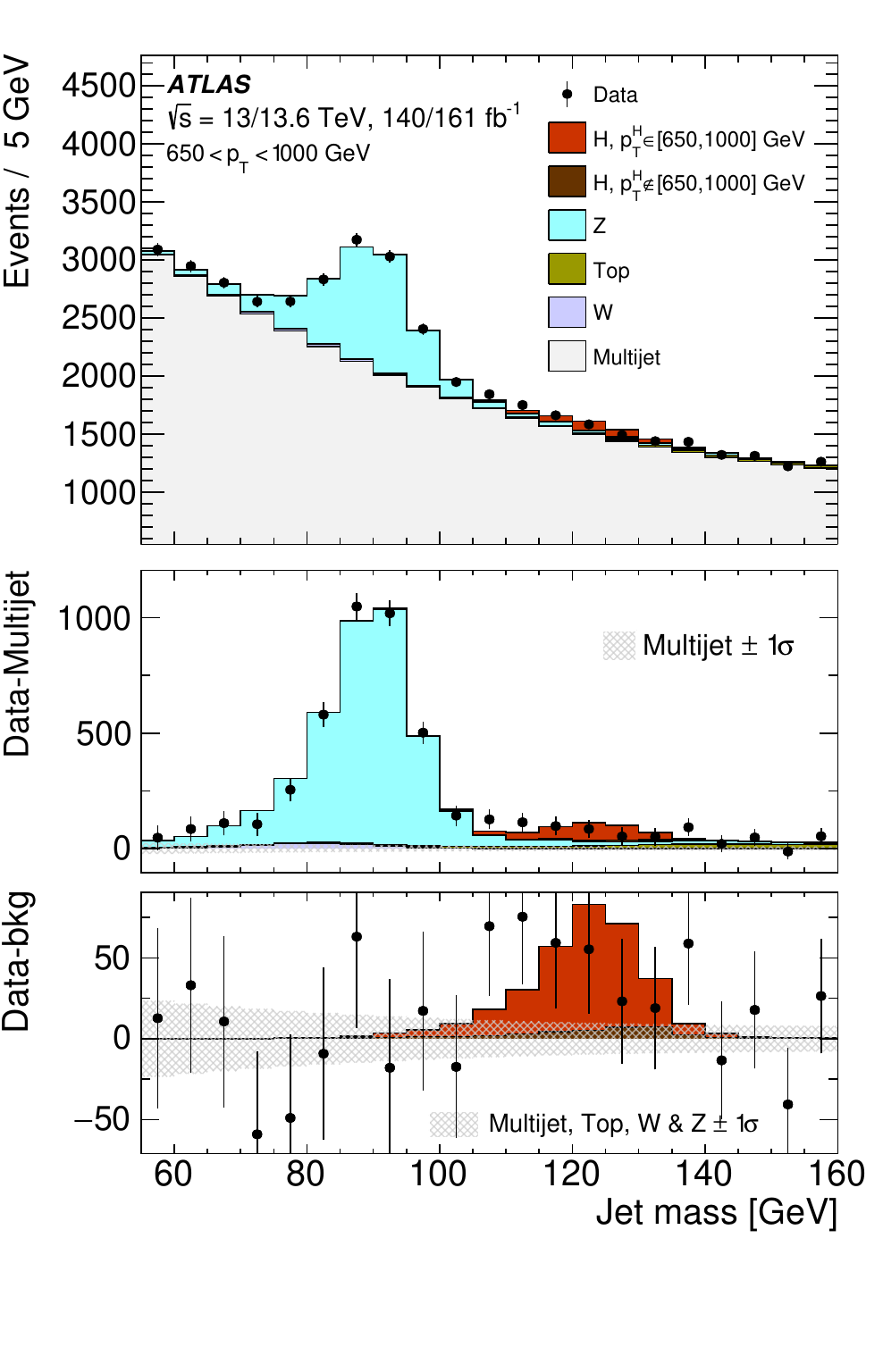}
\vspace*{-1cm}
\caption{}
\end{subfigure}
\begin{subfigure}{0.32\columnwidth}
\includegraphics[width=\textwidth]{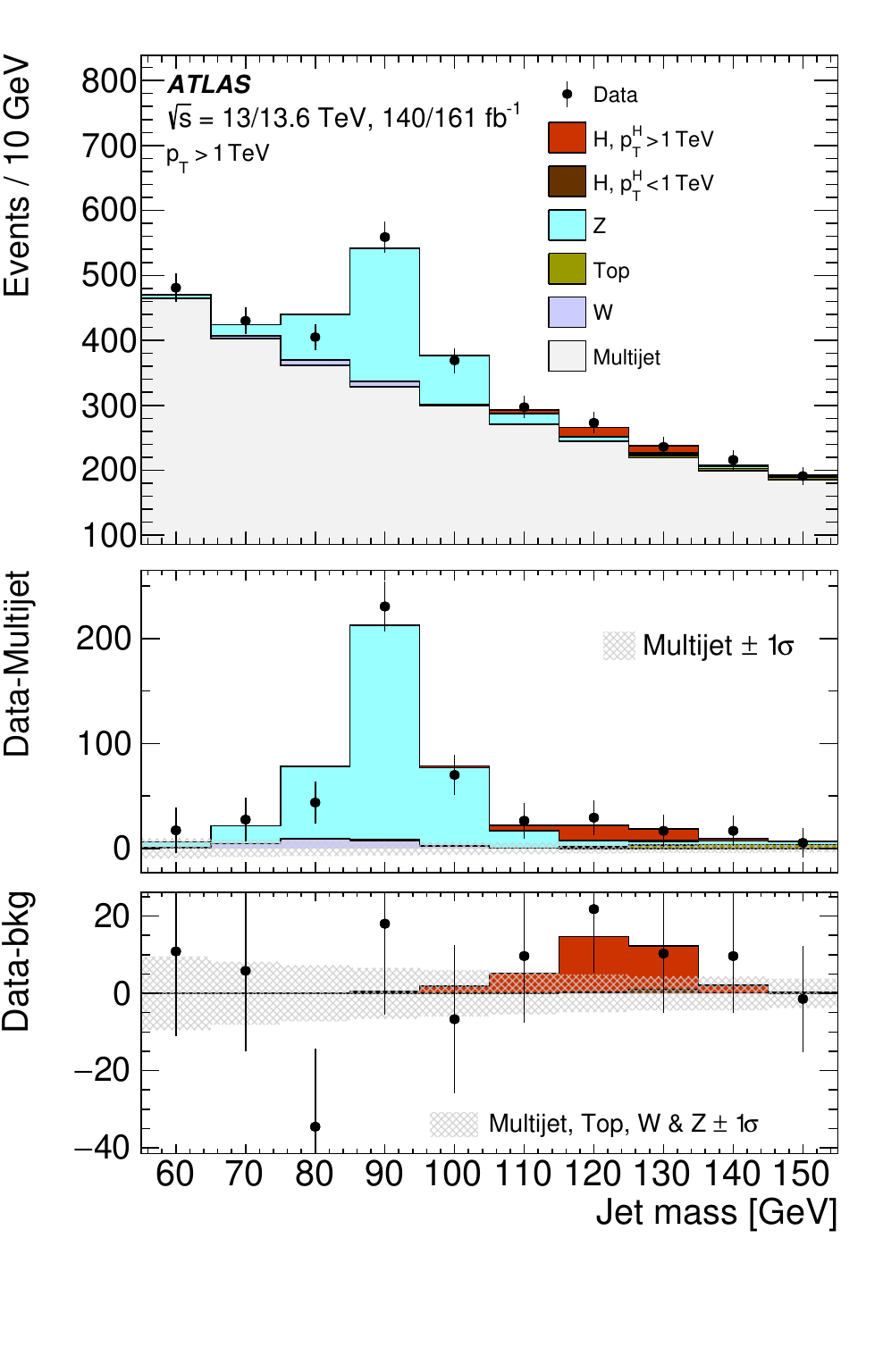}
\vspace*{-1cm}
\caption{}
\end{subfigure}
\caption{
Post-fit jet mass distributions in the (a) $450<\pT<650~\GeV$, (b) $650<\pT<1000~\GeV$, and (c) $\pT>1000~\GeV$ signal regions, summed over jet categories and datasets. The middle panel shows the data with the  multijet background subtracted, and the bottom panel shows the data with all backgrounds subtracted. The fit is performed simultaneously across the twelve signal–control region pairs, with the Higgs boson contribution scaled to the fitted \muH.
\label{fig:DifferentialPostFit} }
\end{figure}

As part of the validation of the analysis strategy and the determination of the GN2X tagging efficiency, the \Zjets event rate is probed in three complementary final states (see \cref{fig:ZvsJetPt}). The \Zmm event rate is extracted from the 96\%-pure \Zmmjets sample with a relative precision of \SIrange{3}{14}{\%}.
The excellent agreement with state-of-the-art theory predictions~\cite{Ridder:2015dxa,Ridder:2016nkl} across the full \pT\ range validates the denominator of \KGNTX.
The inclusive \Vqq event rate derived from fits to the (anti-tagged) VR, which contains 99.8\% of the selected jets, is also consistent.
The uncalibrated \Zbb rate, without \KGNTX applied, extracted simultaneously with the Higgs boson signal strength, is averaged and overlaid for comparison. With increasing \pT, it exhibits a mild upward trend, consistent with residual differences in the GN2X tagging efficiency between data and simulation. This behavior is captured in the in-situ determination of the twelve \KGNTX factors, represented in the lower panel of \cref{fig:ZvsJetPt} after averaging over datasets. For $\pT>1$~TeV, the \Zbb signal is observed
with a significance exceeding five standard deviations from the background-only hypothesis (see ~\cref{fig:DifferentialPostFit}), demonstrating the performance of boosted $bb$ reconstruction and flavor tagging in the TeV regime.

The signal strength for Higgs bosons with reconstructed jet $\pt \ge 450~\GeV$ extracted across all signal regions is
$\muH = 1.53\,\pm 0.27\,\text{(stat.)}\ ^{+0.33}_{-0.27}\,\text{(syst.)}\,\pm 0.17\,\text{(theo.)}$
and corresponds to an observed (expected) significance of $3.8\,(2.5)\,\sigma$ relative to the background-only hypothesis.
Among the experimental uncertainties, the JMR and JMS parameters contribute $\pm0.20$ to the uncertainty on $\mu_H$, while the \KGNTX parameters contribute $\pm0.17$, as determined from the impact method~\cite{Pinto:2023yob}. These are largely statistical in origin, reflecting the finite precision of the \Zbb sample that constrains them, and are expected to decrease with increasing data.
The combined post-fit distributions are shown in \cref{fig:DifferentialPostFit} and exhibit good agreement between the fitted model and the data for all \pT bins.

The Higgs boson yield is also extracted with three independent signal strengths corresponding to intervals of its truth transverse momentum \pTH in a simplified template cross section STXS-like framework~\cite{deFlorian:2016spz,Badger:2016bpw,Amoroso:2020lgh}.
The yield of migrated Higgs boson events with truth $\pTH<450~\GeV$ are constrained to the results from Ref.~\cite{HIGG-2022-04}.
The fitted \muH values are
$1.98\,\pm0.36\,\text{(stat.)}\ ^{+0.53}_{-0.40}\,\text{(syst.)}\,\pm 0.17\,\text{(theo.)}$,
$0.23\,\pm0.67\,\text{(stat.)}\,\pm 0.49\,\text{(syst.)}\,\pm 0.17\,\text{(theo.)}$,
and $2.3\,\pm1.4\,\text{(stat.)}\ ^{+1.3}_{-1.1}\,\text{(syst.)}\,\pm 0.2\,\text{(theo.)}$  %
for \pTH intervals of \SIrange{450}{650}{GeV}, \SIrange{650}{1000}{GeV}, and $>1000~\GeV$, respectively, corresponding to observed (expected) significances of $3.9\, (1.9)\, \sigma$, $0.3\, (1.3) \,\sigma$, and $1.3\, (0.5) \,\sigma$. %
The measurements are compatible with the SM expectations at the 16\% level.
Signal strengths extracted separately in Run~2 and Run~3, and in individual signal regions, are compatible with the reported results, with $p$-values of $\sim$\SIrange{40}{80}{\%}.
The stability of the results has been tested under variations of the fit configuration. Changing the order of $f_N$ by $\pm1$, correlating $\KGNTX$ across runs or jet categories, and decorrelating the JMS and JMR parameters between the $Z$ and $H$ components result in shifts in the measured \muH that are compatible with the reported results.

\begin{figure}[hb!]
\centering
\includegraphics[width=.5\columnwidth]{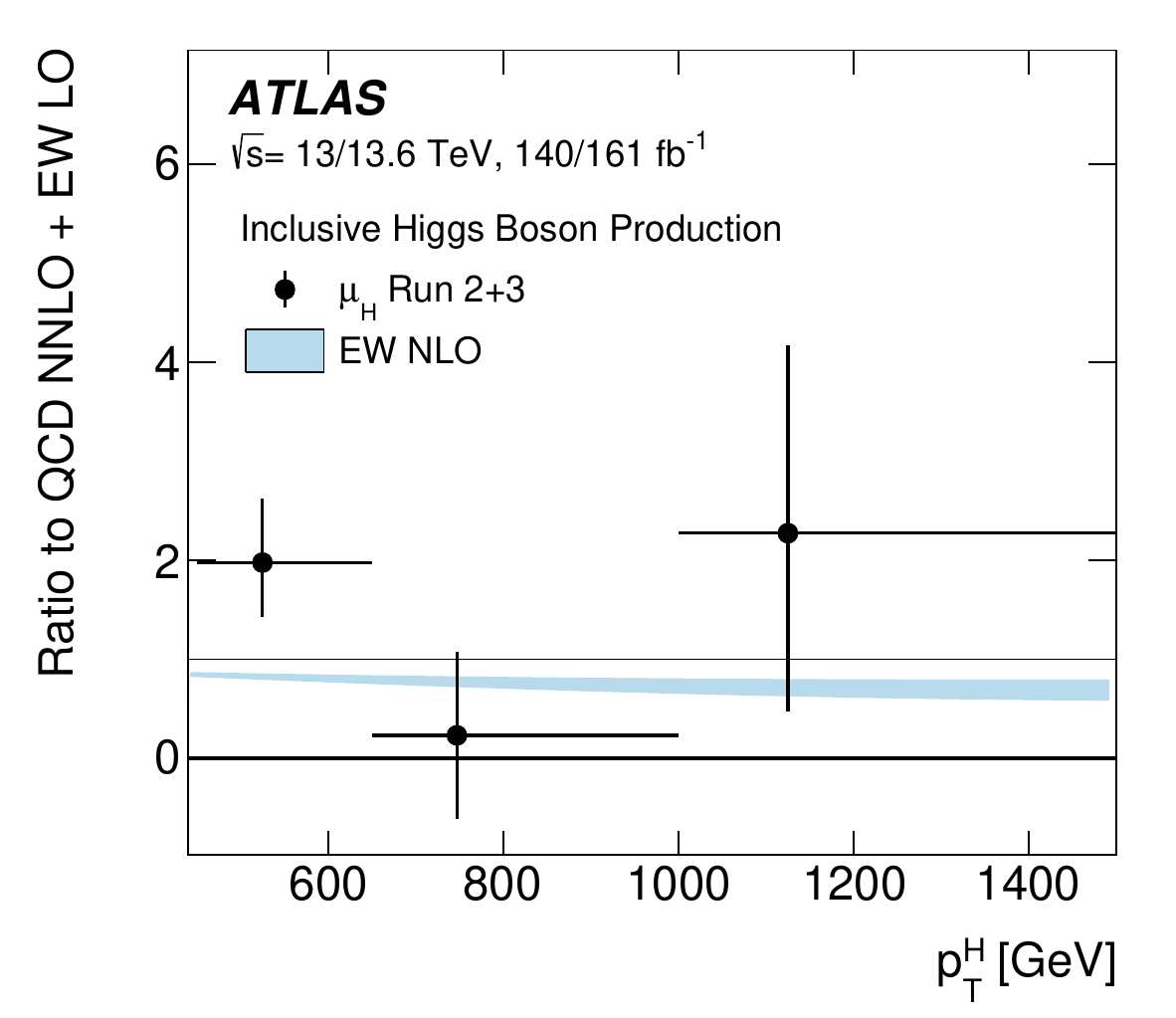}
\caption{Measured inclusive Higgs boson production signal strengths for the Run~2 and Run~3 datasets as a function of \pTH. The EW NLO corrections~\cite{Denner:2014cla}, averaged over the \VH, \VBF and \ttH Higgs production processes, are also shown.
\label{fig:HvsPtH}}
\end{figure}

Measured inclusive Higgs boson production signal strengths for the combined Run~2 and Run~3 dataset are shown in \cref{fig:HvsPtH} as a function of \pTH. The EW NLO corrections derived using the \textsc{HAWK} package~\cite{Denner:2014cla} are shown for the average of the \VH, \VBF ~and \ttH ~Higgs boson production processes, for which they are known, weighted by their acceptance in this analysis.

Separately measuring \muH in Run~2 and Run~3 data with the same fit,
yields $\sigma \times \mathrm{B}(\Hbb)$ values at 13~TeV of
$40^{+21}_{-16}$~fb,
$2.6 \pm 1.7$~fb,
and $1.0^{+1.0}_{-0.8}$~fb
for increasing \pTH intervals of \SIrange{450}{650}{GeV}, \SIrange{650}{1000}{GeV}, and $>1000~\GeV$, respectively.
At 13.6~TeV the corresponding values are
$43^{+22}_{-17}$~fb for \SIrange{450}{650}{GeV},
$<8.7$~fb at 95\% confidence level for \SIrange{650}{1000}{GeV},
and $0.36 \pm 0.72$~fb for $>1000~\GeV$~\cite{hepdata.169580}.

% End of text imported from the .//sections/5.results.tex input file

% The next lines are included from the .//sections/6.conclusion.tex input file
%
%

In conclusion, the ATLAS experiment reports a study of inclusive Higgs boson production at large transverse momentum using the \Hbb decay channel. Evidence is observed for \pT $\ge$ 450~GeV with a signal strength of $1.53\,\pm 0.27\,\text{(stat.)}\ ^{+0.33}_{-0.27}\,\text{(syst.)}\,\pm 0.17\,\text{(theo.)}$, corresponding to an observed (expected) significance of $3.8\,(2.5)\,\sigma$ relative to the background-only hypothesis. Measurements in three \pTH\ intervals are consistent with Standard Model predictions.
This result is enabled by advances in transformer-based jet reconstruction and a simultaneous fit strategy that determines the multijet background from data. It provides the most precise determination to date of Higgs boson production at high transverse momentum and extends sensitivity to potential deviations from the SM at the TeV scale.

% End of text imported from the .//sections/6.conclusion.tex input file

\FloatBarrier
\section*{Acknowledgements}
%

%

% The next lines are included from the .//acknowledgements/Acknowledgements.tex input file
%
%

%
%

We thank CERN for the very successful operation of the LHC and its injectors, as well as the support staff at
CERN and at our institutions worldwide without whom ATLAS could not be operated efficiently.

The crucial computing support from all WLCG partners is acknowledged gratefully, in particular from CERN, the ATLAS Tier-1 facilities at TRIUMF/SFU (Canada), NDGF (Denmark, Norway, Sweden), CC-IN2P3 (France), KIT/GridKA (Germany), INFN-CNAF (Italy), NL-T1 (Netherlands), PIC (Spain), RAL (UK) and BNL (USA), the Tier-2 facilities worldwide and large non-WLCG resource providers. Major contributors of computing resources are listed in Ref.~\cite{ATL-SOFT-PUB-2026-001}.

We gratefully acknowledge the support of ANPCyT, Argentina; YerPhI, Armenia; ARC, Australia; BMWFW and FWF, Austria; ANAS, Azerbaijan; CNPq and FAPESP, Brazil; NSERC, NRC and CFI, Canada; CERN; ANID, Chile; CAS, MOST, NSFC and FRFCU, China; Minciencias, Colombia; MEYS CR, Czech Republic; DNRF and DNSRC, Denmark; IN2P3-CNRS and CEA-DRF/IRFU, France; SRNSFG, Georgia; BMFTR, HGF and MPG, Germany; GSRI, Greece; RGC and Hong Kong SAR, China; ICHEP and Academy of Sciences and Humanities, Israel; INFN, Italy; MEXT and JSPS, Japan; CNRST, Morocco; NWO, Netherlands; RCN, Norway; MNiSW, Poland; FCT, Portugal; MNE/IFA, Romania; MSTDI, Serbia; MSSR, Slovakia; ARIS and MVZI, Slovenia; DSI/NRF, South Africa; MICIU/AEI, Spain; SRC and Wallenberg Foundation, Sweden; SERI, SNSF and Cantons of Bern and Geneva, Switzerland; NSTC, Taipei; TENMAK, T\"urkiye; STFC/UKRI, United Kingdom; DOE and NSF, United States of America.

Individual groups and members have received support from BCKDF, CANARIE, CRC and DRAC, Canada; CERN-CZ, FORTE and PRIMUS, Czech Republic; COST, ERC, ERDF, Horizon 2020 and Marie Sk{\l}odowska-Curie Actions, European Union; Investissements d'Avenir Labex, Investissements d'Avenir Idex and ANR, France; DFG and AvH Foundation, Germany; Herakleitos, Thales and Aristeia programmes co-financed by EU-ESF and the Greek NSRF, Greece; BSF-NSF and MINERVA, Israel; NCN and NAWA, Poland; La Caixa Banking Foundation, CERCA and AGAUR programs from Generalitat de Catalunya and PROMETEO and GenT Programmes Generalitat Valenciana, Spain; G\"{o}ran Gustafssons Stiftelse, Sweden; The Royal Society and Leverhulme Trust, United Kingdom; Eric and Wendy Schmidt Fund for Strategic Innovation, United States of America.

In addition, individual members wish to acknowledge support from Chile: Agencia Nacional de Investigaci\'on y Desarrollo (ANID FONDECYT reg. 1230987, FONDECYT 1230812, FONDECYT 1240864, Fondecyt 3240661, Fondecyt Regular 1240721); China: Fundamental Research Funds for the Central Universities (010-63263105), Chinese Ministry of Science and Technology (MOST-2023YFA1605700, MOST-2023YFA1609300), National Natural Science Foundation of China (NSFC 12275265, NSFC-W2543005); Czech Republic: Czech Science Foundation (GACR - 24-11373S), Ministry of Education Youth and Sports (ERC-CZ-LL2327, FORTE CZ.02.01.01/00/22\_008/0004632), PRIMUS Research Programme (PRIMUS/21/SCI/017); EU: H2020 European Research Council (ERC - 101002463); European Union: European Research Council (BARD No. 101116429, ERC - 101219398, ERC - 948254, ERC 101089007), European Regional Development Fund (HE COFUND GA No.101081355, ERDF), Marie Sklodowska-Curie Actions (GAP-101168829); France: Agence Nationale de la Recherche (ANR-21-CE31-0013, ANR-22-EDIR-0002, ANR-24-CE31-0504-01); Germany: Deutsche Forschungsgemeinschaft (DFG - 469666862); China: Research Grants Council (GRF); Italy: Istituto Nazionale di Fisica Nucleare (LHC-MIUR - 28003/2025), Ministero dell'Università e della Ricerca (NextGenEU 153D23001490006 M4C2.1.1, NextGenEU I53D23000820006 M4C2.1.1, NextGenEU I53D23001490006 M4C2.1.1, SOE2024\_0000023); Japan: Japan Society for the Promotion of Science (JSPS KAKENHI  JP25H0063, JSPS KAKENHI JP22H01227, JSPS KAKENHI JP22H04944, JSPS KAKENHI JP22KK0227, JSPS KAKENHI JP24K23939, JSPS KAKENHI JP24KK0251, JSPS KAKENHI JP25H00650, JSPS KAKENHI JP25H01291, JSPS KAKENHI JP25K01011, JSPS KAKENHI JP25K01023); Poland: Polish National Science Centre (NCN 2021/42/E/ST2/00350, NCN OPUS 2023/51/B/ST2/02507, NCN OPUS nr 2022/47/B/ST2/03059, NCN UMO-2019/34/E/ST2/00393, UMO-2022/47/O/ST2/00148, UMO-2023/49/B/ST2/04085, UMO-2023/51/B/ST2/00920, UMO-2024/53/N/ST2/00869); Spain: Agència de Gestió d'Ajuts Universitaris i de Recerca. (AGAUR - 2023 BP 00141), Ministry of Science and Innovation (RYC2019-028510-I, RYC2020-030254-I, RYC2021-031273-I, RYC2022-038164-I), Ministerio de Ciencia, Innovación y Universidades/Agencia Estatal de Investigaci\'on (EU NextGenerationEU (PRTR-C17.I1), PID2022-142604OB-C22); Sweden: Carl Trygger Foundation (Carl Trygger Foundation CTS 22:2312), Swedish Research Council (Swedish Research Council 2023-04654, VR 2021-03651, VR 2022-03845, VR 2022-04683, VR 2023-03403, VR 2024-05451, VR 2025-05940), Knut and Alice Wallenberg Foundation (KAW 2023.0366); Switzerland: Swiss National Science Foundation (SNSF - PCEFP2\_194658); United Kingdom: The Binks Trust, Royal Society (NIF-R1-231091); United States of America: U.S. Department of Energy (ECA DE-AC02-76SF00515), John Templeton Foundation (John Templeton Foundation 63206), Neubauer Family Foundation.

%
%

% End of text imported from the .//acknowledgements/Acknowledgements.tex input file

%
%

%
\clearpage
\appendix
\part*{End Matter}
\addcontentsline{toc}{part}{End Matter}

% The next lines are included from the .//sections/7.end_matter.tex input file
%

\section{Analysis selection}

Candidate jets are organized into regions according to the GN2X tagger output. The SRs are populated with leading or subleading candidate jets passing the GN2X tagger requirements. Anti-tagged jets populate either the CR or the VR. Jets are sorted into the various regions according to criteria given in \cref{tab:jetsorting}.

\begin{table}[h]
\centering
\caption{Assignment of events to analysis signal or control/validation regions based on the GN2X tagging decisions of the leading and subleading candidate jets. The symbol `--' indicates the jet is not included in the jet mass distribution of any region.}
\begin{tabular}{cc|cc}
\multicolumn{2}{c|}{GN2X tagging decision} & \multicolumn{2}{c}{Jet's assigned region} \\
Leading jet & Subleading jet & Leading jet & Subleading jet \\
\hline
Pass  & Pass & SRL &  -- \\
Pass  & Fail & SRL & CRS/VRS \\
Fail  & Pass & CRL/VRL & SRS \\
Fail  & Fail & CRL/VRL & CRS/VRS \\
\end{tabular}
\label{tab:jetsorting}
\end{table}

\section{Higgs boson sample generation and theory uncertainty}
Higgs boson \ggF\ production was simulated at next-to-leading-order (NLO) QCD accuracy with finite mass effects by using the {\textsc Hj-MiNLO} ~\cite{Hamilton:2015nsa,Hamilton:2012np,Campbell:2012am,Hamilton:2012rf} prescription with the \POWHEGBOX program~\cite{Alioli:2010xd,Nason:2004rx,Frixione:2007vw} and \NNPDF[3.0nnlo]~\cite{Ball:2014uwa} parton distribution function (PDF).
NLO QCD accuracy for \VBF\ and \ttH\ production and LO accuracy for $gg\to\VH$ (where $V$ is a $W$ or $Z$ boson) production was achieved using the \POWHEGBOX~\cite{Nason:2009ai,Hartanto:2015uka,Alioli:2010xd,Nason:2004rx,Frixione:2007vw} program and \NNPDF[3.0nlo]~\cite{Ball:2014uwa} PDF.
The Higgs boson production cross sections were scaled to match predictions at next-to-next-to-leading order (NNLO) in QCD using the heavy-top effective field theory combined with exact NLO top quark mass effects for \ggF\ production, NNLO for \VBF\ production, and NLO for \VH\ and \ttH\ production~\cite{Becker:2020rjp}.
The reweighting is performed in bins of Higgs boson transverse momentum using scale factors defined as the ratio of the higher-order calculation values to the corresponding generator predictions.
The resulting corrections are typically at the \SIrange{5}{10}{\%} level for \ggF\ and \VBF, at the few-percent level for \ttH, and up to $\sim10\%$ for \VH\ at high \pT.
As differential predictions at $\sqrt{s}=13.6$~TeV are not yet available at the same perturbative accuracy, the \pT-dependent scale factors derived at 13~TeV are applied to the Run~3 samples.
Corrections for EW NLO effects are derived using the \textsc{HAWK} package~\cite{Denner:2014cla} for \VH, \VBF ~and \ttH ~Higgs production processes.

The theoretical cross section uncertainties are taken from Ref.~\cite{Becker:2020rjp}. The accuracy for the different production processes, known at different order in QCD, varies considerably. They are included conservatively applying the scale variation uncertainty estimated for the approximate NNLO calculation of the \ggF\ cross section to all Higgs boson production modes (11\%). Top-quark mass scheme uncertainties, estimated at 10--18\% for the \ggF\ process~\cite{ATL-PHYS-PUB-2023-031}, are included.  In addition the PDF uncertainty for \ggF\ is derived comparing to the alternative \PDFforLHC[15nlo] PDF set~\cite{Butterworth:2015oua}. A flat 10\% difference is observed in \pTH and \mJ. Acceptance effects from the variation of the \pTH distribution from a seven-point scale variation is estimated and found to vary with \pTH between 2 and 5\%. No substantial differences are observed in the mass shape when jet shower systematics are evaluated by comparing \Pythia to \Herwig. These uncertainties are propagated to the total signal yield according to the relative contribution of each production mode, resulting in an overall uncertainty of approximately 17--18\% depending on \pT.

\section{Multijet modeling}
The modeling of the multijet background was validated using simulated QCD samples produced with different generators and fragmentation models,  namely \PYTHIA[8] with the Lund string fragmentation model~\cite{Andersson:1983ia,Sjostrand:1984ic}, \HERWIG[7]~\cite{Bellm:2017jjp} with the cluster~\cite{Winter:2003tt} and Lund fragmentation models, \SHERPA with the native Ahadic fragmentation model~\cite{Winter:2003tt} and \POWHEG+\PYTHIA[8]. These studies show generator-dependent differences in the shape of the tagged and anti-tagged jet mass spectra, driven primarily by variations in heavy-flavor composition and parton-shower modeling. The ratio of the normalized SR and CR spectra is well described by a second order function in all tested models, supporting the robustness of the parametrization used in the fit to data.

\section{Systematic uncertainties and stability tests}

The correlation between the fitted \muH and the \Zbb rates, determing the \KGNTX factors, is found to be smaller than $20\%$ across all regions.

The in-situ determinations of the simulation-to-data $bb$ tagging efficiency correction factors \KGNTX and the jet mass scale (JMS) and resolution (JMR) factors require the transfer of results on jet tagging and mass performance from the \Zboson to the Higgs. This assumes coherence in response and simulation modeling between the two resonant states.

Several checks were performed to ensure the validity of this procedure. Possible generator-related differences between \Hbb (\POWHEG+\PYTHIA) and \Zbb (\SHERPA) production are evaluated through MC-to-MC comparisons.

Variations of the \SHERPA fragmentation model change from the native Ahadic to the Lund string model induce a change of the \Zbb tagging efficiency by about 5\%, which is small compared to the uncertainty from the GN2X data-to-simulation efficiency correction factors ($\pm0.17$ on $\mu_H$). No additional uncertainty is assigned.

Compatible JMS and JMR behavior for $\pT>450~\GeV$ is seen in studies of boosted \Wqq jets from \ttbar events generated with \POWHEG+\PYTHIA and \SHERPA. As a cross-check, the JMS and JMR parameters are decorrelated between the \Zbb and \Hbb templates in the fit; the resulting \muH values remain consistent with the baseline result within uncertainties.
The JES and JER uncertainties are derived from measurements in \Zboson $+$ jet, \Zmm events dominated by light-flavor jets and applied to the $b$-tagged jets used in this analysis. The bJR response correction is consistent between simulated \Zqq and \Zbb events, indicating that the regression is largely flavor independent, as expected from its training dataset including all jet flavors in multijet samples. %

The full set of systematic uncertainties is summarized in \cref{tab:systematics}.

\begin{table*}[!htbp]
\begin{center}
\caption{Summary of the systematic uncertainties used in the fit model. Ranges in the magnitude column describe the envelope of uncertainty size across \pT, where larger uncertainties usually affect the highest \pT\ bin. Magnitudes for shape-only variations represent effective up and down variations of the involved quantity, i.e.\ the Jet mass resolution (scale) changes the peak width (position) by the relative amount shown in the table.
The uncertainty contribution on the Higgs boson signal strength from the given source of systematic uncertainty is noted in the last column labeled $\delta \muH$ as calculated with the impact method~\cite{Pinto:2023yob}.
\label{tab:systematics}}
\begin{tabular}{l c cc l c}
\hline
\multirow{2}{*}{\footnotesize{Systematic}}
& \multirow{2}{*}{\footnotesize{Magnitude ($\%$)}}
& \multicolumn{2}{c}{\footnotesize{Effect on observable}}
& \multirow{2}{*}{\footnotesize{Processes concerned}}
& \footnotesize{$\delta \mu_{H}$} \\
&
& \footnotesize{Norm.}
& \footnotesize{Shape}
& \\
\hline
\multicolumn{6}{c}{\footnotesize{Experimental systematics}} \\
\hline
SR Jet Mass Resolution (JMR) & 25 & \xmark & \cmark & \Zbb, \Hbb & \\
SR Jet Mass Scale (JMS) & 5 &  \xmark & \cmark  & \Zbb, \Hbb & \\
CR Jet Mass Resolution & 25 &  \xmark & \cmark  & \Vjets & \\
CR Jet Mass Scale & 5 &  \xmark & \cmark  & \Vjets & \\
JMR+JMS               &    &         &         &        & $\pm 0.19$ \\ \hline
\Zmm normalization & 3--14 & \cmark & \xmark & \Zbb &  \\
\KGNTX & 5--30 & \cmark & \xmark & $Z\to b\bar{b}$, $H\to b\bar{b}$ & $\pm 0.17$\\
\pT resolution & $1-7$ & \cmark & \xmark & $H\to b\bar{b}$ & $\pm 0.02$\\
Trigger efficiency & $2$ & \cmark & \xmark & all MC & $\pm 0.02$\\
\ttbar normalization & 25 & \cmark & \xmark & \ttbar & $\pm 0.05$\\
CR \Vjets normalization & 10 & \cmark & \xmark & $V=W+Z\to q\bar{q}$ & $\pm 0.07$ \\
Higgs spurious signal & 0.1--0.3 \sigmafit &  \cmark & \xmark  & $H\to b\bar{b}$  & $\pm 0.15$\\
\hline
\multicolumn{6}{c}{\footnotesize{Theoretical systematics}} \\
\hline
Higgs cross section & 11 &  \cmark & \xmark  & $H\to b\bar{b}$ & \\
top-quark mass scheme & 10--18 &  \cmark & \xmark  & $\ggF\to H$ & \\
Higgs PDFs           & 10 &  \cmark & \xmark  & $H\to b\bar{b}$ & \\
Higgs scale variations (acc. only) & 2-5 &  \cmark & \xmark  & $H\to b\bar{b}$ & \\
Higgs theory uncert.               &     &         &         &                 &
$\pm 0.17-0.18$ \\
\hline
\end{tabular}
\end{center}
\end{table*}

As an additional consistency check, \muH has been determined for the Run~2 and Run~3 datasets separately. The result is $\muH=1.89\pm0.65~(1.28\pm0.49)$ for the Run~2(3) datasets, where the quoted uncertainty accounts for statistics and experimental systematics.

\FloatBarrier

% End of text imported from the .//sections/7.end_matter.tex input file

%

%
%

%
%
\printbibliography
\clearpage
% ATLAS Collaboration author list
% Reference date of HIGP-2024-01 is 2026-02-23
% Author list last updated on date 01-JUL-26
% Data extracted on 01-Jul-2026 for paper reference HIGP-2024-01
% at 5:24pm
 
\begin{flushleft}
\hypersetup{urlcolor=black}
{\Large The ATLAS Collaboration}

\bigskip

\AtlasOrcid[0000-0002-6665-4934]{G.~Aad}$^\textrm{\scriptsize 102}$,
\AtlasOrcid[0000-0001-7616-1554]{E.~Aakvaag}$^\textrm{\scriptsize 17}$,
\AtlasOrcid[0000-0002-5888-2734]{B.~Abbott}$^\textrm{\scriptsize 121}$,
\AtlasOrcid[0000-0002-0287-5869]{S.~Abdelhameed}$^\textrm{\scriptsize 83b}$,
\AtlasOrcid[0000-0002-1002-1652]{K.~Abeling}$^\textrm{\scriptsize 54}$,
\AtlasOrcid[0000-0001-5763-2760]{N.J.~Abicht}$^\textrm{\scriptsize 48}$,
\AtlasOrcid[0000-0002-8496-9294]{S.H.~Abidi}$^\textrm{\scriptsize 30}$,
\AtlasOrcid[0009-0003-6578-220X]{M.~Aboelela}$^\textrm{\scriptsize 44}$,
\AtlasOrcid[0000-0002-9987-2292]{A.~Aboulhorma}$^\textrm{\scriptsize 36e}$,
\AtlasOrcid[0000-0001-5329-6640]{H.~Abramowicz}$^\textrm{\scriptsize 154}$
% ...
\bigskip
\\

% ...
$^{17}$Department for Physics and Technology, University of Bergen, Bergen; Norway.\\
% ...
$^{30}$Physics Department, Brookhaven National Laboratory, Upton NY; United States of America.\\
% ...
$^{36}$$^{(a)}$Facult\'e des Sciences Ain Chock, Universit\'e Hassan II de Casablanca;$^{(b)}$Facult\'{e} des Sciences, Universit\'{e} Ibn-Tofail, K\'{e}nitra;$^{(c)}$Facult\'e des Sciences Semlalia, Universit\'e Cadi Ayyad, LPHEA-Marrakech;$^{(d)}$LPMR, Facult\'e des Sciences, Universit\'e Mohamed Premier, Oujda;$^{(e)}$Facult\'e des sciences, Universit\'e Mohammed V, Rabat;$^{(f)}$Institute of Applied Physics, Mohammed VI Polytechnic University, Ben Guerir; Morocco.\\
% ...
$^{44}$Physics Department, Southern Methodist University, Dallas TX; United States of America.\\
% ...
$^{48}$Fakult\"{a}t Physik, Technische Universit{\"a}t Dortmund, Dortmund; Germany.\\
% ...
$^{54}$II. Physikalisches Institut, Georg-August-Universit\"{a}t G\"ottingen, G\"ottingen; Germany.\\
% ...
$^{83}$$^{(a)}$Khalifa University of Science and Technology, Abu Dhabi;$^{(b)}$New York University Abu Dhabi, Abu Dhabi;$^{(c)}$United Arab Emirates University, Al Ain;$^{(d)}$University of Sharjah, Sharjah; United Arab Emirates.\\
% ...
$^{102}$CPPM, Aix-Marseille Universit\'e, CNRS/IN2P3, Marseille; France.\\
% ...
$^{121}$Homer L. Dodge Department of Physics and Astronomy, University of Oklahoma, Norman OK; United States of America.\\
% ...
$^{154}$Raymond and Beverly Sackler School of Physics and Astronomy, Tel Aviv University, Tel Aviv; Israel.\\
% ...
$^{a}$ Also at Affiliated with an institute formerly covered by a cooperation agreement with CERN.\\
$^{b}$ Also at An-Najah National University, Nablus; Palestine.\\
$^{c}$ Also at Borough of Manhattan Community College, City University of New York, New York NY; United States of America.\\
$^{d}$ Also at Center for Interdisciplinary Research and Innovation (CIRI-AUTH), Thessaloniki; Greece.\\
$^{e}$ Also at Centre of Physics of the Universities of Minho and Porto (CF-UM-UP); Portugal.\\
$^{f}$ Also at CERN, Geneva; Switzerland.\\
% ...
$^{*}$ Deceased

\end{flushleft}

% Created with Glance <Atlas.Glance@cern.ch>

%

%
%
%
%
%
%

%
%
%
%
%
%

\end{document}